\documentclass[referee]{an}
\usepackage{graphicx}
\usepackage{times}
\usepackage{fancyhdr}
\usepackage{natbib}
\begin{document}
 
\title{Weak redshift discretisation in the Local Group of Galaxies?}
 
\author{W{\l}odzimierz God{\l}owski\inst{1}, Katarzyna Bajan\inst{2}, Piotr Flin\inst{3}}
 
\institute{Astronomical Observatory of the Jagiellonian
University, ul. Orla 171, 30-244 Krakow, Poland \and the Henryk
Niewodniczanski Institute of Nuclear Physics, Polish Academy of
Sciences, 31-342 Krakow, ul. Radzikowskiego 152,  Poland \and
Institute of Physics, Pedagogical University, ul. Swietokrzyska
15,  25-406 Kielce, Poland}
 
\date{Received $<$date$>$;
accepted $<$date$>$;
published online $<$date$>$}
 
\abstract {We discuss the distribution of radial velocities of
galaxies belonging to the Local Group. Two independent samples of
galaxies as well as several methods of reduction from the
heliocentric to the galactocentric radial velocities are
explored. We applied the power spectrum analysis using the Hann
function as a weighting method, together with the jackknife error
estimation. We performed a detailed analysis of this approach. The
distribution of galaxy redshifts seems to be non-random. An excess
of galaxies with radial velocities of $\sim 24 \, \mbox{km} \cdot
\mbox{s}^{-1}$ and $\sim 36 \, \mbox{km} \cdot \mbox{s}^{-1}$ is
detected, but the effect is statistically weak. Only one peak for
radial velocities of $\sim 24 \, \mbox{km} \cdot \mbox{s}^{-1}$
seems to be confirmed at the confidence level of 95\%.
}
 
\authorrunning{God{\l}owski et al.}
 
\titlerunning{Redshift discretisation in LG}

\correspondence{godlows@oa.uj.edu.pl}
 
\maketitle
 
\section{Introduction}
 
In the large-scale Universe, one aspect of the search for
regularities is connected by testing the radial velocity
of galaxies. These velocities can be observed as having arbitrary
values, regular patterns regarded as periodisation, discretisation
or quantization of galaxy redshifts. The discretisation of
redshift for astronomical objects can be discussed independently
for three cases, namely galaxies, quasars and large-scale
periodicity (120 Mpc). The latter studies have been discussed
(Bajan et al. 2003), so we will not repeat these
investigations here. In our previous paper (Bajan et al. 2004),
the quest for quasar redshift periodicity is described. Here,
only the main points from its history, together with some recent
results, will be mentioned.
 
The story of redshift periodisation started when Burbidge (1968)
noted the existence of sharp peaks in redshift distribution,
grouped around the values of $z=0.01$ and $z=1.95$. He also found
periodicity in the redshift distribution, which can be described
by the formula $z=0.01\cdot n$.
 
The existence of periodicity in the $\log(1+z)$ scale was established by
Karlsson (1971). In early stage of these investigations, association
between bright low-redshift galaxies and quasars was sought in
order to check for the non-cosmological origin of QSO redshifts,
as well as to test for the possibility of ejection of QSO from parent
galaxies, a problem investigated even now (Bell 2004).
The link between GRBs and massive star formation is also
suggested. Over the next years, the effect of periodisation was
either confirmed or denied on the basis of incorrectly applied
statistics and/or possible selection effect. The selection effects
have been discussed by many authors.  Some of them, (e.g.  Karlsson 1971),
claimed that the observed redshift distribution is not due to the
selection  effects, while other authors (e.g. Basu 2005) expressed
the opposite opinion.
 
Burbidge and Napier (2001) took into account a new sample of high
redshift quasars located in projection on celestial spheare 
close $(\leq 10^{''})$ to the low redshift
galaxies. The existence of periodicity in the $\log (1+z)$ scale,
was confirmed. The 2dF QSO Redshift Survey containing over 10000
objects and the 2dF Galaxy Redshift Survey with over 100000
galaxies served as an objective basis in the periodicity search
for quasar-galaxy pairs. Having investigated 1647 objects,
Hawkins, Maddox \& Merrifield (2002) found no periodicity. Bell
(2004) used two quasar samples, one with high redshifts of
$2.4-4.8$ and the other with low redshifts of $0.02-0.2$. He
showed that all peaks in these two redshift distributions occur at
the previously predicted preferred values.
 
Napier \& Burbidge (2001) re-examined the redshift distribution
in the 2dF QSO Redshift Survey, detecting periodicity. Arp, Roscoe
\& Fulton (2005) showed that the maxima in redshift distribution
in the 2dF and DSS surveys fit well into the formula. Bell's
(2004) investigation showed clear periodicity in the redshift
distribution of QSO. The latest work on redshift periodicity was
done by Basu (2005). He investigated the same 33 objects (GRBs,
QSO and active galaxy) as Burbidge (2003) and contrary to
Burbidge, he claimed that all existing peaks are due to
observational and analytical selection effects.

The discussion of galaxy redshift quantization started with the
work of Tifft (1976). He claimed that the redshifts of galaxies in
the Coma cluster were preferentially offset from each other in
multiples of $72.46 \, \mbox{km} \cdot \mbox{s}^{-1}$. A few years
later, the existence of global periodicity was reported (Tifft \& Cocke 1984);
however, the period was not $72 \, \mbox{km} \cdot
\mbox{s}^{-1}$, but $36 \, \mbox{km} \cdot \mbox{s}^{-1}$ or
possibly $24 \, \mbox{km} \cdot \mbox{s}^{-1}$. It should also be
noted that the values of $\sim 24 \,\mbox{km} \cdot
\mbox{s}^{-1}$, as reported by Tifft \& Cocke (1984), were  not
confirmed by the later investigations.
 
We know that this subject is not popular and usually very
suspicious at first glance. However, on the basis of the claimed
results of redshift periodisation, several theoretical papers
pointed out the necessity for so-called new physics were
published.
 
Therefore, we decided to check if the discretisation does occur.
We share the opinion expressed by Hawkins et al. (2002) that all these
effects should be carefully checked. They claimed:

``The criticism usually leveled at this kind of study is that the
samples of redshifts have tended to be rather small and selected in a
heterogeneous manner, which makes it hard to assess their
significance.  The more cynical critics also point out that the
results tend to come from a relatively small group of astronomers who
have a strong prejudice in favour of detecting such unconventional
phenomena.  This small group of astronomers, not unreasonably,
responds by pointing out that adherents to the conventional
cosmological paradigm have at least as strong a prejudice towards
denying such results.
 
We have attempted to carry out this analysis without prejudice.
Indeed, we would have been happy with either outcome: if
the periodicity were detected, then there would be some fascinating new
astrophysics for us to explore; if it were not detected, then we would
have the reassurance that our existing work on redshift surveys, etc,
has not been based on false premises.''

Iwanowska (1989) claimed that the spatial distribution of galaxies
belonging to the Local Group and  globular clusters located close
to our Galaxy is linear, i.e., these objects form long chains. For
galaxies located in such lines, Zabierowski (1993) distinguished 5
different redshift groups finding that the observed  periodicity
is consistent with the Tifft's (1976) value of $72 \, \mbox{km}  \cdot
\mbox{s}^{-1}$. He counted mean velocities in each group, using
old data (RC3 (de Vaucouleurs et al. 1991). Rudnicki, God{\l}owski
\& Magdziarz (2001) addressed the same problem, using better observational
data. They considered 40 galaxies as well as globular clusters and
performed a simple statistical analysis based on the calculation
of the mean values of redshifts in bins and their dispersions.
They tested strict quantization, that is precise multiplication of
the value of $36 \, \mbox{km} \cdot \mbox{s}^{-1}$, finding no
effect. However, with the result of the power spectrum analysis,
the weak effect of periodisation, i.e. the grouping around some
values of galaxy radial velocities, was noted for galaxies
situated in two of the Iwanowska's lines.

The expected interpretation of galaxy redshift quantisation is
that redshifts are not cosmological. Moreover, there are
suggestions that clusters of quasars evolved into clusters of
galaxies, so the galaxy redshift distribution should reflect this
fact.
 
Tifft (1996) claimed that galaxy redshift distribution is quantised,
because time is quantised. If this effect global, it means that 
it should be observed in all galaxy structures.
 
The number of galaxies in the Local Group is small but in the
present work we consider all galaxies regarded as members of the
Local Group (hereafter referred to as the LG) (Irwin, 2000; van
den Bergh 1999). We applied the power spectrum analysis as the
statistical tool in our investigation.
 
We used the standard method, i.e. power spectrum analysis,
corrected for its possible disadvantages when applied to this
problem. Moreover, we adopted a totally new approach. We use the
Hann function as a weighting factor together with the jackknife
error estimation, first used by Hawkins et al. (2002)
(who investigated quasars, not galaxies). Because it is quite
a new approach, we are enclosing several plots showing not only the
results but also detailed analysing properties of the method. Such
detailed analysis was not performed in the Hawkins et al. (2002)
paper. Our result is that a weak effect of
periodisation is observed for galactocentric velocities, while for
heliocentric and LG-centric radial velocities no effect was
observed.
 
The paper is organized in the following manner. The second section
presents observational data, and the third section describes the
method of analysis applied. The fourth section consists of our
results. In the last section we give our conclusions.

\section{Observational Data}
 
There is no common agreement as to which galaxies belong to the
dynamical aggregate called the Local Group (LG). We considered 55
objects (see Tables 1 and 2) in our vicinity taken from the
Irwin's list (Irwin 2000) (based on Mateo (1998)) together with 7
galaxies, mostly within the Maffei group, which could also {\it
probably} be regarded as the LG members (Iwanowska, 1989) (marked
in Tables 1 and 2 as $M$). Considering various parameters, van den
Bergh (1999) concluded that only 32 objects can be LG members,
while 3 further objects can be regarded as possible LG members. We
decided to perform all calculations using these two sets of data.
Sets based on the Irwin List were denoted as $A$, while those
based on the van den Bergh's (1999) were denoted as $B$.
 
It should be noted that in the van den Bergh list there are 7
galaxies (not including Phoenix) without redshift, while the
Irwin's list contains 2 more such objects (Cetus and Cam A). In
this manner, 46 and  28 galaxies remain to be analysed in each
sample respectively. We denoted these samples as I and II.
Separately, we analysed pure Irwin data (39 galaxies) as sample
III.
 
The search for any systematic effects in the distribution of
redshifts requires precise knowledge of redshifts. There are some
discrepancies among various redshift determinations. In order to
avoid the influence of this factor on the results, we analysed
separately samples with radial velocities taken from the Irwin
list (2000) (sample I) from those from van den Bergh's (1999)
(sample II). Whenever no data was available for sample I, we
decided to use the data from RC3 (de Vaucouleurs et al. 1991).
 
The latest data allows us to find redshifts for 8 of the total 9
galaxies (marked as $N$ in Tables 1 and 2), these redshifts were not
known prior to these analysis. Redshifts for
7 of them are noted by Karachentsev et al. (2002). We repeated the
analysis with these objects considered. However, we decided to
exclude Tucana because its radial velocity measurement is
uncertain. It should be noted that for the 6 remaining galaxies
redshifts are also included in the new version of the Irwin's list
(sample IV). Moreover, for two galaxies, the new version of the
Irwin's list replaces the old redshift (NGC 147 $157\, \mbox{km}
\cdot \mbox{s}^{-1}$ and NGC 221 $190\, \mbox{km} \cdot
\mbox{s}^{-1}$ ). In our analysis we do not include, Cam A either
(noted in NED as uncertain, with no errors determined). Finally,
adding 6 galaxies to the Irwin's and the van den Bergh's lists we
obtain samples of 45 galaxies (sample IV) and 34 galaxies (sample
V) respectively. A detailed list of galaxies is given in Tables 1
and 2. Column (10) gives heliocentric radial velocity for set $A$
(samples I, III and IV), while column (11) for  set $B$ (samples
II and V). As seen from the data, errors of measuring radial
velocities are really small for the majority of galaxies.
 
The heliocentric radial velocities of galaxies should be corrected
for the motion of the Sun relative to the center of our Galaxy
and/or LG. This is usually done by applying the correction to the
center of our Galaxy only. There are several prescriptions for how
to perform this reduction. In this paper, we applied  various
galactocentric corrections known from literature. We analysed the
following solar motions:
 
\noindent a): velocity $v=232 \,\mbox{km} \cdot \mbox{s}^{-1}$ in
the direction of $l=88^o$, $b=2^o$ as proposed by \citet{9}
denoted as $a$,
 
\noindent b): or $v=233 \pm 7 \,\mbox{km} \cdot \mbox{s}^{-1}$ in
the direction of $l=93^o \pm 10^o$, $b=2^o \pm 5^o$ also proposed
by \citet{10} denoted as $b$,
 
\noindent c): and $v= 213 \pm 10 \,\mbox{km} \cdot \mbox{s}^{-1}$
in the direction of $l=93^o \pm 3^o$, $b=2^o \pm 5^o$ according to
\citet{11} denoted as $c$.
 
There was also a suggestion that the local standard  of rest has
an  additional radial component directed outward our Galaxy (Clube
\& Waddington 1989). We decided to add this component into
correction $a$. In such a way, we obtained a new correction
denoted as $d$ with the following values: $v=234 \,\mbox{km} \cdot
\mbox{s}^{-1}$, $l=98^o$, $b=2^o$ (Guthrie \& Napier 1991).
Pure heliocentric data that is without any corrections hereafter
denoted as $e$ were analysed, too.
 
Moreover, the correction of the Sun's motion with respect to the
LG centroid was included (Coutreau \& van den Bergh 1999).
This gave the velocity of the motion of the Sun relative to the
LG of $v= 306 \pm 18 \,\mbox{km} \cdot \mbox{s}^{-1}$ toward an
apex at $l= 99^o \pm 5^o$ and $b= -3.5^o \pm 4^o$ ($f$). Two
additional corrections for the motion of the Sun relative to the
LG were also taken into account. The correction denoted as $g$
with the value of $v=305 \pm 136 \,\mbox{km} \cdot \mbox{s}^{-1}$
toward $l=94^o \pm 48^o$, $b=-34^o \pm 29$ calculated (Rauzy \&
Gurzadyan 1998) from two subgroups (ours and that of M31) in the
LG, and "the historical" velocity (Yahil, Tammann \& Sandage
1977), denoted as $h$: $v= 308 \pm 23 \,\mbox{km} \cdot
\mbox{s}^{-1}$ directed toward $l=105^o \pm 5$ and $b=-7^o \pm
4^o$ were used.
 
It is interesting to note that the spatial structure of the Galaxy
is flat. Therefore, the correction to galactocentric radial
velocities lies in the Galactic plane. Therefore, we checked
how the situation changes when a fictitious perpendicular
component of the Sun velocity vector is assumed. We considered a
velocity of $v= 224 \,\mbox{km} \cdot \mbox{s}^{-1}$ directed
toward $l= 109^o$ and $b= 65^o$ as correction $i$. Finally, the
new proposed correction of the Sun's motion (Karachentsev et al.
2002) in respect to the LG of $v= 316 \, \mbox{km} \cdot
\mbox{s}^{-1}$ toward the apex at $l= 93^o$ and $b= -4^o$, denoted
as $j$, was used. All of the corrections are summarized in Table
3.

\section{Method of analysis}
 
The power spectrum analysis (PSA) (Yu \& Peebles 1969; Webster
1976; Guthrie \& Napier 1990), together with the Rayleigh
test (Mardia 1972; Batschelet 1981), has been used as a
statistical tool. It was shown (Newman, Haynes \& Terzian 1994)
that this method is very useful for finding periodicity among
irregularly distributed points. The Rayleigh's test (Mardia,
1972; Batschelet 1981) is a simple test of uniformity which
allows one to detect periodicities in irregularly distributed
points. For a given frequency, the Rayleigh power spectrum
corresponds to the Fourier power spectrum, as well as measuring
the probability of the existence of a sinusoidal component.
 
Let us assume that $m$ points are distributed along a finite line with
coordinates $x_i$, where i$=1, \ldots, m$.
We can define the phase with respect to the period $P$:
    $$
\Phi_i = 2 \pi x_i /P. \eqno(1)
    $$
The phase $\Phi_i$ which corresponds to the $i$-th $x$ coordinate
unambiguously describes the radius vector $I_i$.
 
The length of the vector $R$, constituting the sum of all $I_i$
vectors, can be used for testing  isotropy of the distribution of
points $x_i$ with respect to the period $P$
    $$
R(P)  = \sum_{i=1}^N I_i(P). \eqno(2)
    $$
From the formal point of view, the vector $R$ describes a random
walk in the plane after $m$ steps.
 
We use the statistic $R^2$ defined by:
   $$
 R^2(P) = \left(\sum_{i=1}^N \cos \Phi_i\right)^2 + \left(\sum_{i=1}^N \sin \Phi_i\right)^2, \eqno(3)
   $$
The probability distribution of $R^2$ can be calculated from the null
hypothesis:
   $$
p_i d\Phi = d\Phi/ 2 \pi \qquad \qquad \Phi  \in (0, 2\pi), \eqno(4)
   $$
where $p_i (\Phi)d\Phi$ is the probability that phase $\Phi_i$
corresponding to point $x_i$ is located in the interval
$(\Phi,\Phi + d\Phi)$. The distribution of $R^2$ corresponds to
the Fourier power spectrum  for function $f(\Phi) = \sum \delta
(\Phi - \Phi_i)$.
 
It is known that the variable:
    $$
  s(P)= 2R^2/m, \eqno(5)
    $$
has the following distribution (Webster 1976):
    $$
p(s,m)ds = ds (m/4) \int_{0}^\infty
J_o^m(\omega)J_o(\omega\sqrt{ms/2})\omega d\omega, \eqno(6)
    $$
where $J_o$ is a Bessel function. This formula is also valid for a
small number of objects, which is the case of the LG.
 
The distribution $p(s,m)$ is calculated numerically by integrating
approximations of the Bessel function using the Romberg method
(Press et al. 1992). For large $m$, it could be also aproximated as a
$\chi^2$ distribution with 2 degrees of freedom.
 
Error bars of $s(P)$ can be estimated using the "jackknife"
technique of drawing all possible samples of $N-1$ values from $N$
data points, repeating the power spectrum analysis on these
samplings. Such a procedure allowed us to calculate the standard
deviation in the derived values of $s$ $\sigma_j(P)$. The best
estimator for the standard errors in the value of $s$ is then just
$\sqrt{N-1}\sigma_j$ (Hawkins et al. 2002).
 
The simulation of the power spectrum for random uniformly
distributed data is presented in Figure 1. The diagrams showing
the values of $s$-statistics versus $1/P$ present the results of
the power spectrum analysis. There are several peaks in each
diagram. These peaks allow one to find each possible period, as
well as to investigate the significance of each particular peak in
the power spectrum. The level of significance of  each peak is
given by $C =1-(1 - p)^{n_{\mathrm{t}}}$, where $p$ is the probability of
obtaining, from the theoretical, random distribution the value of
the $s$-statistics equal to or greater than the observed value of
the $s$-statistics, while $n_{\mathrm{t}}$ is the number of independent peaks
within the analysed frequency range (Lake \& Roeder 1972;
Guthrie \& Napier 1991).
 
The additional test increasing the efficiency of the test for weak
clustering, following (Webster 1976; Scott 1991), is based
on the summation over the whole power spectrum. This sum gives the
value of $SI=\sum s_i$, with a $\chi^2$ distribution with $2n_{\mathrm{t}}$
degrees of freedom. Thus, the expected value of the $SI$ statistic
is $2n_{\mathrm{t}}$. The $SI$-test can be used for testing the randomness
of the distribution.
 
The clustering statistics $Q$ is equal to the value of the
$SI$-statistics over its expected value. The expected value for
$Q$ statistics is calculated in the case of a random walk
($2n_t$). For a random distribution, the expected value of $Q$ is
equal to 1, with the error of: $\sigma (Q) = 1 / \sqrt{n_{\mathrm{t}}}$
(Webster 1976). We tested the hypothesis that the value of Q is
greater than unity rather than equal to this value. In our
analysis, we decided to consider the first 50 peaks. In this case,
at the significance level of 0.05, the critical value of the $SI$
statistics is 124.3, with 135.8 for the significance level of 0.01,
also note that $\sigma$ (Q) = 0.14.
 
Newman et al. (1994) pointed out that Yu and Peebles'
version of PSA can be applied correctly only when a uniform
distribution function is tested. As seen from Figure 2, such an
approximation could be accepted for raw data but not when a
correction for the solar motion is included.
 
Hawkins et al. (2002) discussed the power spectrum
method when the data is not uniformly distributed. They proposed to
use the window function and showed that the power spectrum method
works well in that case. Now, the power of $s$ at the period of
$P$ is given via the formulae (Hawkins et al. 2002):

$$
s(P)= 2R^2/\sum_{i=1}^N w_i^2, \eqno(7)
$$
where
 
$$
R^2(P) = (\sum_{i=1}^N w_i\cos \Phi_i)^2 + (\sum_{i=1}^N w_i\sin \Phi_i)^2, \eqno(8)
$$
 
Following Hawkins et al. (2002), we repeat our
analysis using the Hann's function as a weight:
   $$
w_i= \frac{1}{2} \left[ 1- \cos \left( \frac{2\pi x_i}{L} \right) \right]
   $$
where $L$ is chosen to cover the whole range over which the data
is selected.
 
It should be pointed that now the expected value of the clustering
statistics $Q$ is not necessarily equal to 1, especially for a
small number of points. Therefore, we decided to test how the PSA
with the Hann's weighting function is working. The observed
distribution of galactocentric redshift in the LG is normal (Fig.
2). Using the K-S tests, we checked that the distributions denoted
as I and II and containing 46 and 28 galaxies respectively, are
normal at the significance level of $\alpha = 0.05$. Thus, in the
future analysis we use the Gaussian redshift distribution, as
theoretically expected distribution.
 
We run 1000 simulations for data distributed normally, and with
variance taken from the real data with correction $a$. We find
that in the case of sample I (46 galaxies), the mean value of $Q$
is equal to 1, but $\sigma (Q) = 0.195$. For sample II (28
galaxies), the mean value of $Q$ decreases and is equal to 0.90,
but $\sigma (Q) = 0.177$. This means that the noise value of $s=2$
(see Hawkins et al. 2002) is slightly changed.
The example of the simulation of the power spectrum for normally
distributed data weighted using the Hann's function is presented
in Figure 3. One can see that when the data are apodized, the
expected height of the peak decreases.
 
\section{Result}
 
We started our investigations with a classical power spectrum
analysis (Yu \& Peebles 1969; Webster 1976). The two basic
samples of galaxies (based on sets $A$ and $B$), containing 46 and
28 objects each, denoted as sample I and II respectively, were
analysed. For each sample, 10 different reductions of heliocentric
radial velocities to galactocentric velocities or a LG centroid,
were performed. In Figure 4 the power spectra of sample I
analysed as $s$ versus $1/P$ diagrams in the range of $1/P \in
(0.0025, 0.05)$ is shown. The visual impression is that some peaks are
clearly distinguishable, which can be regarded as concentration
around some particular values of the period. These values are
close to $36 \,\mbox{km} \cdot \mbox{s}^{-1}$, and $24 \,\mbox{km}
\cdot \mbox{s}^{-1}$. However, this impression is not confirmed by
statistical analysis.

The result of  our analysis  is given  in Table 4 (sample I) (and
Table 5 for sample II respectively). The first column of both
tables describes the analysed sample, while the next two give the
values of the velocity $v$ at which the most prominent peak was
observed, together  with the probability $P(s)$ that the peak is
generated by a random distribution (when we restrict for period $P
\in (20, 100)\,\mbox{km} \cdot \mbox{s}^{-1}$).
 
From Table 4, it is easy to see that the most prominent peaks are
statistically not significant, at the significance level of 95\%.
Therefore, we concluded that in random distribution such peaks can
also appear. The power spectrum analysis shows
that the peaks observed in the period distributions are consistent
with the assumption of randomness.
 
The second aim of the power spectrum analysis was to check for the
possibility of weak clustering using $SI$ and $Q$-statistics.
The result of  our analysis  is also given in Table 4. The last
two columns of the table contain the values of $SI$ and $Q$
statistics. It can be easily seen that in the case of
galactocentric correction prescriptions $a$, $b$, $c$, both $SI$
and $Q$-statistics confirm the hypothesis that the redshift
distributions are non-random at the confidence level of 95\%. For
prescriptions $a$ and $b$ this seems to be confirmed even at the
level of 99\%.
 
For the prescription $d$ computed using the probable velocity of
the local rest frame, we do not find any departure from isotropy.
A comparison with the previous cases allows us to state that this
vector does not correspond to reality. The reduction to the LG
centroid, using any of our vectors, did not reveal any
statistically significant individual preferred values of
velocities and showed that the distributions are random.
 
The high value of $s$ given by the PSA is expected at scales
greater than the clustering scale (Yu \& Peebles  1969; Scott 1991).
We deal with one structure (Local Group) only, so this is
not the cause of the reported non-randomness. Nevertheless, we
noted that from the theoretical point of view the non-random
distribution detection by the PSA can be due to the inclusion of
some external objects. In order to examine this possibility, we
performed three tests:
 
i) we consider sample II, from which all objects with untypical
values of redshift were excluded (van den Bergh 1999),
 
ii) we repeat our analysis of sample I without the Argo galaxy,
which is its most pronounced external member,
 
iii) because the artificially high values of $s$ given by the PSA
are observed mainly in the first mode (Webster 1976), for the
period we perform the PSA also in the interval $i=2,\ldots,51$
instead of the first 50 peaks (this is also the reason why we
restrict our analysis to $P \in (20, 100)\,\mbox{km} \cdot
\mbox{s}^{-1}$).
 
All these changes alter the statistics values, but for the
prescriptions $a$ and $b$, the PSA distribution remains
non-random.

Karachentsev et al. (2002) showed that, for galaxies belonging to
the Local Group, errors in heliocentric velocities are generally
$\pm 5\, \mbox{km} \cdot \mbox{s}^{-1}$ or smaller. This is true
for galaxies belonging to sample II. However, it should be
noted that some of the galaxies belonging to sample I have large
errors in their reported radial velocities. This is the case for
the sample of 7 galaxies (marked in  Tables 1 and 2 as $M$) which
Iwanowska (1989) regarded as the {\it probable} LG members
(although this is still very questionable). It should be pointed
out that large errors in the data can mask any existing
periodisation (however, this cannot be a source of a fictitious
signal in PSA). Now we repeated our analysis for the pure Irwin's
list, withouth those 7 galaxies (sample III). The result (Table 5)
is similar to that obtained for sample I but the value of
statistics $SI$  for reduction $a$, $b$ and $c$ is even larger than
in sample I. This confirms that these 7 galaxies denoted as $M$ in
Tables 1 and 2 should be rejected from further analysis.
 
When we add 6 new galaxies (marked in Tables 1 and 2 as $N$), we
obtain (for samples IV and V analysed now) that the PSA
distribution clearly remains non-random for prescriptions $a$
only; however a weak effect survives for prescriptions $b$ and
$c$ as well (see Table 5).
 
Still, to be sure that the obtained effect is real, we should take
into account the fact that our data are not uniformly distributed.
In that case (Hawkins et al. 2002), the PSA must
take into account the data window function. Therefore, we repeated
our analysis using the Hann's function as a weighting function
(Hawkins et al. 2002). For sample II, we clearly obtain
that the distribution of the PSA remains non-random for reduction
$a$ (Table 6, statistic $Q$). Figure 5 also shows that, for
prescriptions $c$, the most prominent peak is significant at about
$2 \sigma$ level above the noise value. For prescriptions $a$ and
$b$, the significance  of the most prominent peak is reduced below
the $2 \sigma$ level. A similar situation is seen in Figure 5 for
sample V, where for prescriptions $a$ and $c$ the peak is
significant at about $2 \sigma$ level. It should be noted that the
PSA with an appropriate window function does not show any signs of
periodicity in other cases. This suggests significant
periodisation close to $24 \, \mbox{km}\cdot \mbox{s}^{-1}$.

Moreover, it should be pointed out that using the window function
we change not only the expected value of the clustering statistics
$Q$, but also the expected height of the peak. As a result, the
probability $P(s)$ formally obtained on the basis of Equation 6
(presented in Table 6) is not valid any longer. Again we run 1000
simulations for the normally distributed data with  variance taken
from the real data. The histogram for the distribution of the
height of the most significant peak with data weighted using the
Hann's function for sample II (28 galaxies) with  variance taken
from the real data with correction $a$, is presented in the left
panel of Figure 6, the cumulated distribution is presented
in the right panel. We can see that peaks with $2R^2/m \ge 8.2$
are significant at the 95\% level. Please note that, at
least for prescription $c$, for samples II and V the most
pronounced peaks are significant at the level of 95\%. A similar
situation occurs for sample V, prescription $a$. This confirms
weak periodisation close to $24 \,\mbox{km} \cdot \mbox{s}^{-1}$
and is in contrast with the previous cases (without using the data
window function), where all peaks were not significant.

\section{Conclusions}
 
In our analysis, we considered the nearby galaxies which are
widely accepted to be LG members. Moreover, we defined a sample of
galaxies according to several physical properties and not only to
spatial distribution, as belonging to the LG. We took into
account several possible ways of correcting the heliocentric
velocity to the galactocentric velocity. The correction to the LG
centroid was also considered. We used the best data currently
available.
 
The preliminary statistical analysis excluded the possibility of
strict redshift quantization in the LG (Rudnicki et al. 2001).
The lack of strict multiplicity of  the value of $36 \,\mbox{km}
\cdot \mbox{s}^{-1}$ based on good data does not confirm
Zabierowski's claim (Zabierowski 1993) to this effect.

We obtained that distributions of galaxy redshift seem to be
non-random when the correction for the motion of the Sun relative
to the center of our Galaxy is taken into account. The argument
revealing a possible significance of the result is the fact that
it is practically independent of the samples. There is some
periodisation though which in some cases is close to $36
\,\mbox{km} \cdot \mbox{s}^{-1}$, while in  other cases it is
close to $24 \,\mbox{km} \cdot \mbox{s}^{-1}$. However, this
effect is statistically weak. The statistical significance of our
result as seen from the Tables is rather low.
 
The most important argument is that the result survives the PSA
when the data window function is taken into account. This provides
that the effect found is real. However, only periodisation close
to $24 \,\mbox{km} \cdot \mbox{s}^{-1}$ seems to be confirmed at
the significance level of 95\%. One can see that the strongest
effect is obtained for galaxies from the van den Bergh's list. Van
den Bergh (1999) concluded that only 32 objects can be LG members,
while 3 further objects can be regarded as possible LG members.
The Irwin's list contains more galaxies for whom membership in
the LG is questionable.
 
With correction for the Sun's motion with respect to the LG
centroid, we obtained no periodisation. We did not obtain any
statistically significant preferred values of velocities in this
case. It is interesting to note that the spatial structure of the
Galaxy is flat. Therefore, the correction to the galactocentric
radial velocities lies in the galactic plane. We also
checked how the situation changes when a fictitious perpendicular
component of the Sun's velocity vector is assumed. In this case
there is no periodisation. This result, together with the analysis
with prescription $e$ (pure heliocentric velocity) shows that the
effect is diluted for an artificial value of this vector. Thus,
this provides evidence that the observed weak periodisation is not
an artifact of the galactocentric correction.

The latest data allows us to include redshifts for 6 galaxies
determined by Karachentsev et al (2002). When we repeated the
analysis with these objects, the effect decreased.
However, it should be pointed out that these objects are connected
with M31 (and they are probably M31 satelites). Thus, any possible
periodisation could be stronger than for other galaxies masked
by the local influence.
 
The interpretation of this result is neither unique nor clear. We
would like to point at the following possibilities:
 
1. The effect of periodisation of galaxies in the LG can be
connected artificially with the manner of galactocentric
correction due to either velocity determination or to the data
itself but the above-mentioned diminishing of non-randomness
when artificial velocities are used does not seem to support
this conclusion.

2. Periodisation is found only when galactocentric velocities are 
included, and it is not connected with the LG centroid. This is 
possible, for example, when the effect is global as claimed by Tifft
and it is not connected with the LG.
 
3. Periodisation is a relic of quantum effects in the early
Universe (Tifft 1996). Thus, the effect is not connected with the
LG, which was formed during much later epoch.
 
4. Periodisation is a real phenomenon, also connected with the LG,
but the correct LG structure is not known at present.

Clearly, power spectrum analysis is a good tool for studying
non-randomness in velocity data. The main result of the paper is
that the distributions of galaxy redshift seem to be non-random,
but each specific individual particular value of periodisation is
statistically marginal. Only periodisation close to $24
\,\mbox{km} \cdot \mbox{s}^{-1}$ seems to be confirmed, and only
at the significance level of 95\%. One of the reasons is that
errors in the data are large. However, it should be noted that in
the version of the PSA considered, non-systematic errors can not
produce "false periodisation" but only "destroy" any real existing
effect. Measurement errors will be important if explicitly taken
into account. In such a version of PSA, we analysed the sample
from the point of view of the objects likelihood rather than
using its best values. However, such a procedure changes the initial
assumptions of the method as well as the analysed statistics, therefore
it is not considered in our present paper. It will be analysed in our
future work. Further investigation should concern such large-scale
structures as the Local Supercluster, Coma/A1367, the Perseus or
Hercules Superclusters, when more accurate HI velocity data become
available.

\section*{Acknowledgments}
 
WG still remembers and owes much to fruitful discussions with late
Pawe{\l} Magdziarz. We thank Professor Church for discussions.
This paper has used  the NASA/IPAC Extragalactic Data (NED),
operated by the Jet Propulsion Laboratory, California Institute of
Technology, with the National Aeronautic and Space Administration.

\twocolumn
 
\begin{figure}
\vskip 6cm
\includegraphics{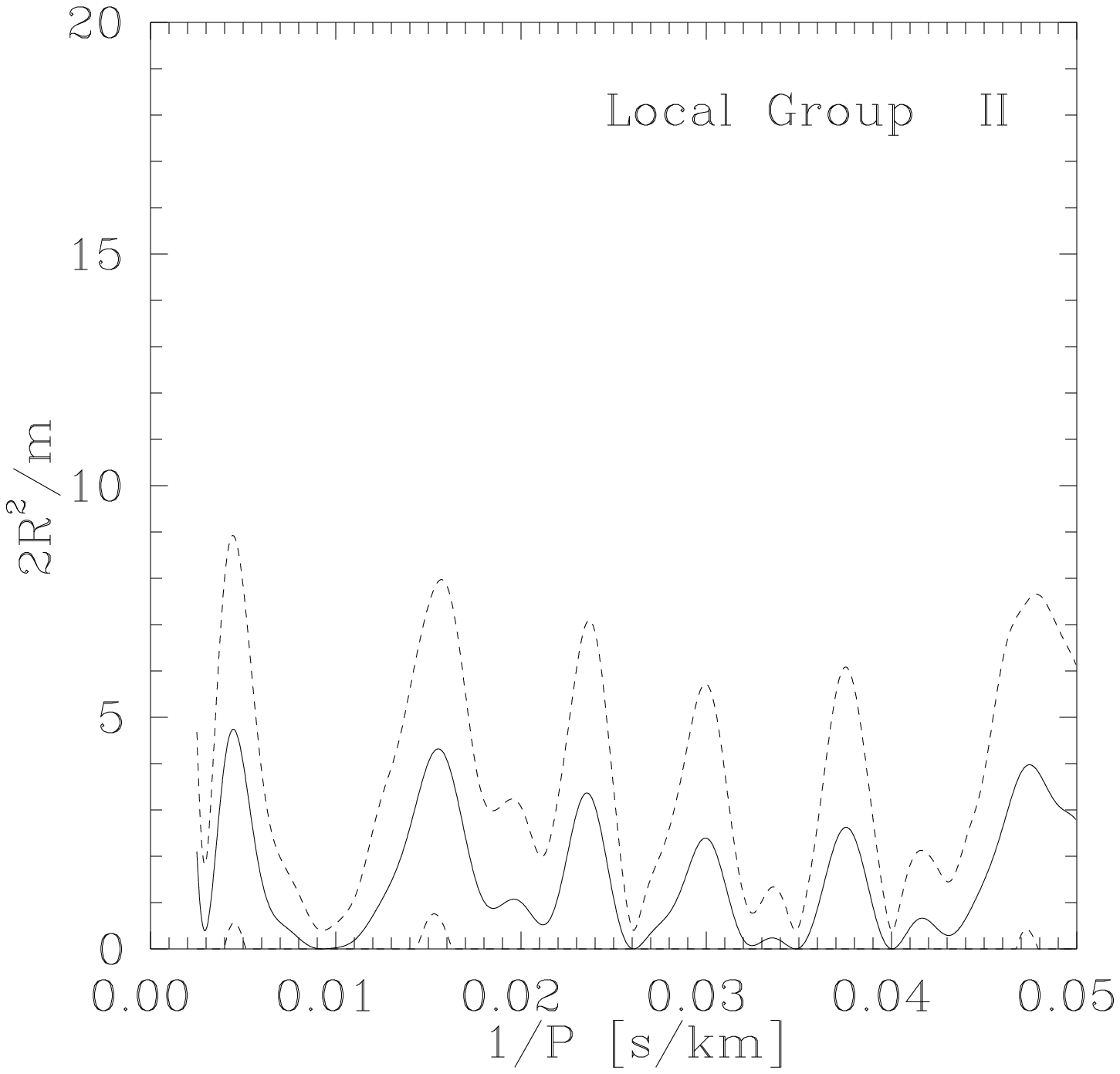}
\caption{Example of simulation of the raw power spectrum of a random uniform
distribution --- sample II (28 galaxies). Dashed lines show errors
derived by  applying the jackknife estimator. \label{fig1}}
\end{figure}
 
\begin{figure*}
\vskip 10cm
\includegraphics{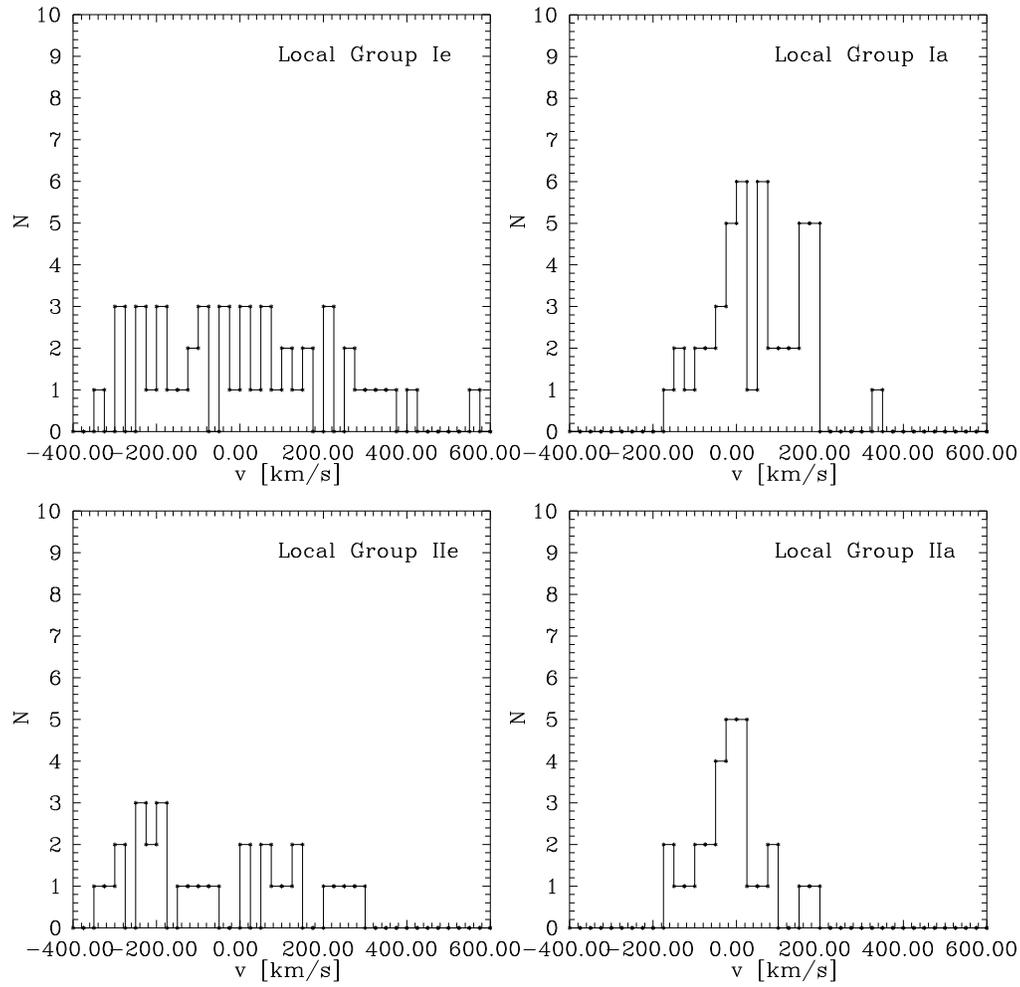}
\caption{Redshift distribution histogram  for  raw data (left side), and those
corrected for the solar motion (correction $a$). \label{fig2}}
\end{figure*}

\begin{figure}
\vskip 6cm
\includegraphics{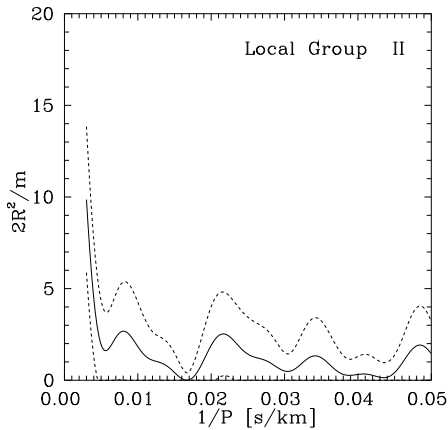}
\caption{Example of simulation of the power spectrum for  normal
distribution derived with  data weighted using the Hann's function ---
sample II (28 galaxies). Dashed lines show errors
derived by  applying the jackknife estimator. \label{fig3}}
\end{figure}

\begin{figure*}
\vskip 20cm
\includegraphics{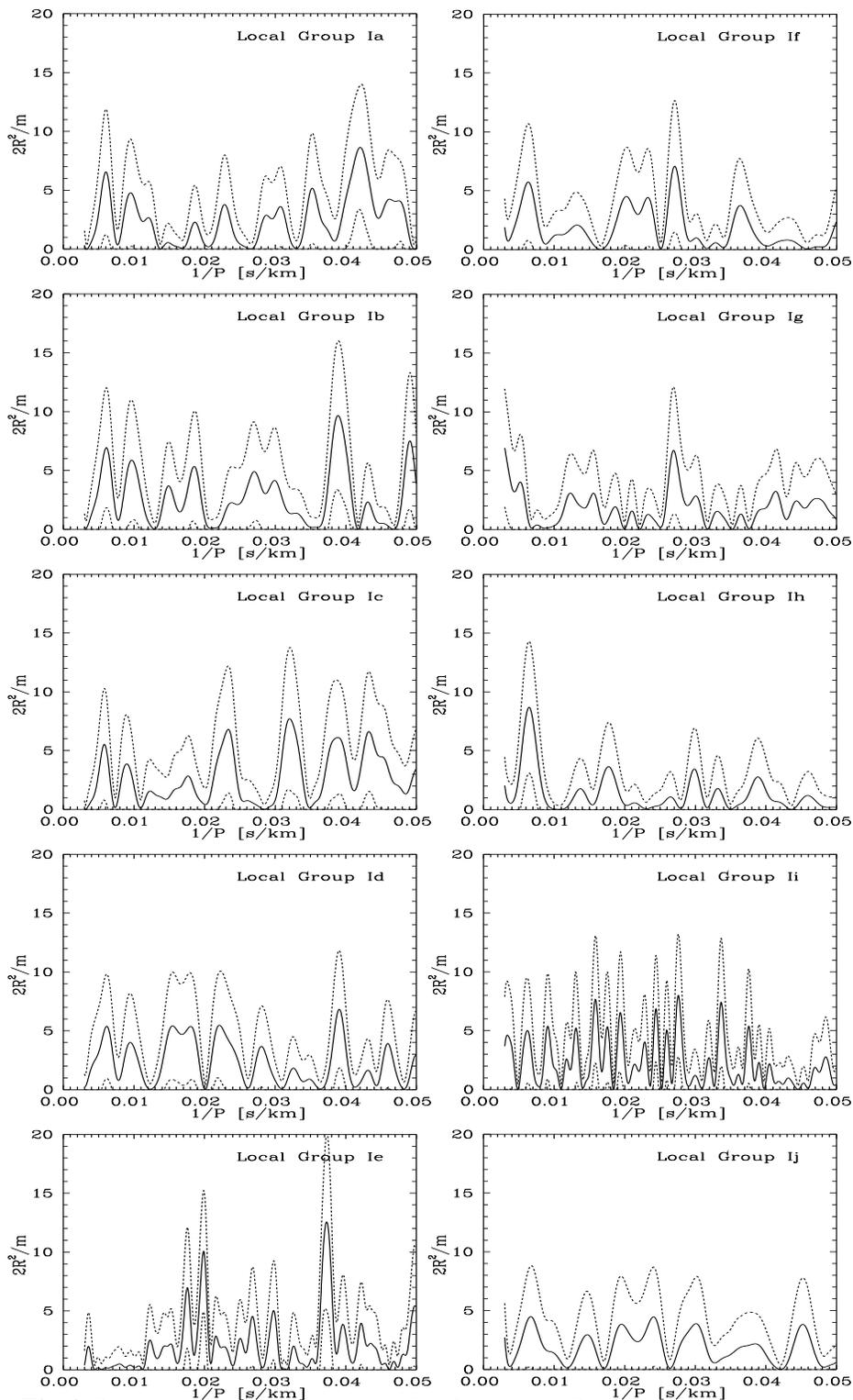}
\caption{Raw power spectrum of sample I (46 galaxies). Dashed lines show
errors  derived by applying the jackknife estimator.\label{fig4}}
\end{figure*}

\begin{figure*}
\vskip 20cm
\includegraphics{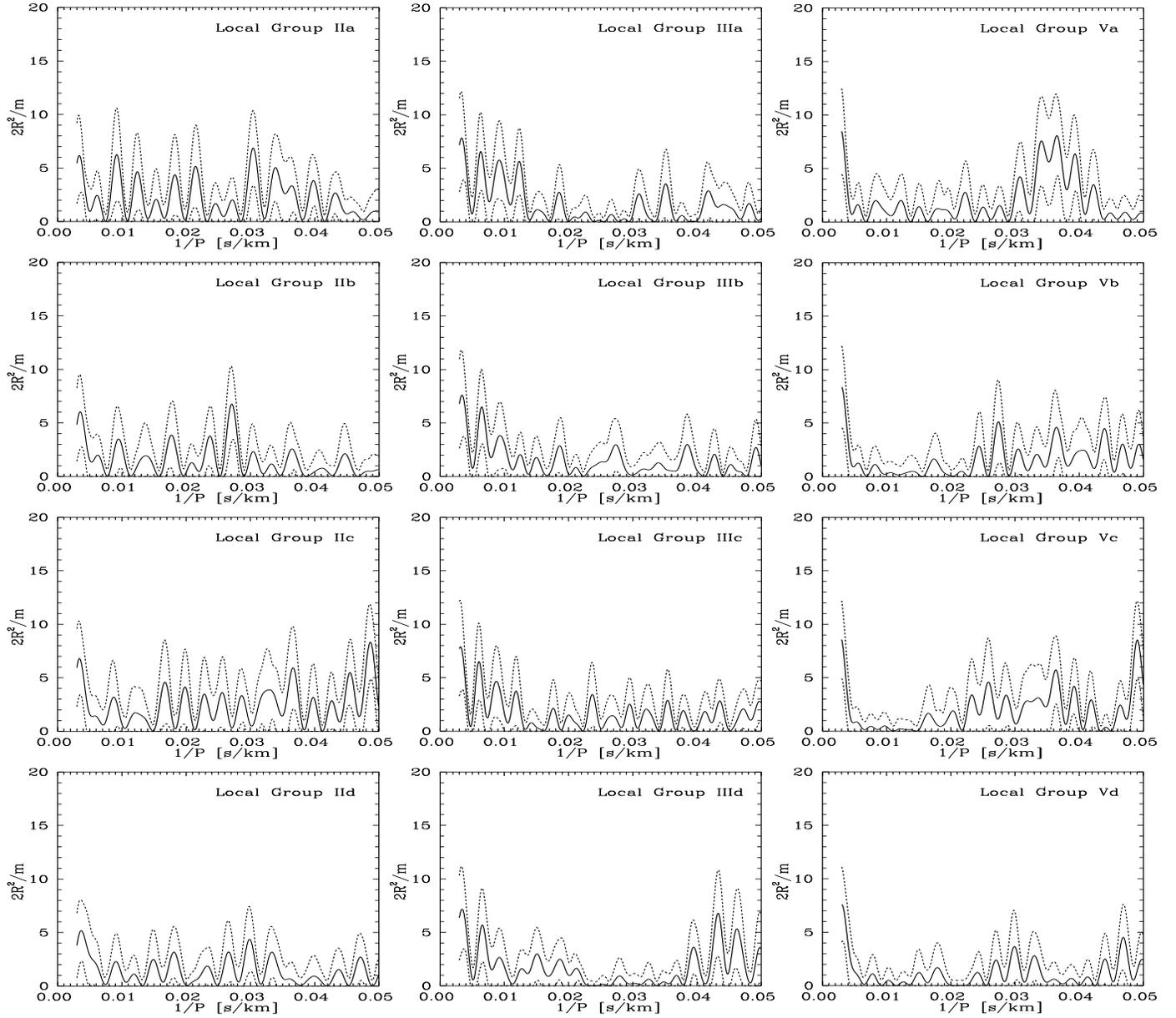}
\caption{Power spectrum of sample II (28 galaxies), III (39 galaxies)
and V (34 galaxies) after apodization with the Hann's function .\label{fig5}}
\end{figure*}

\begin{figure*}
\vskip 6cm
\includegraphics{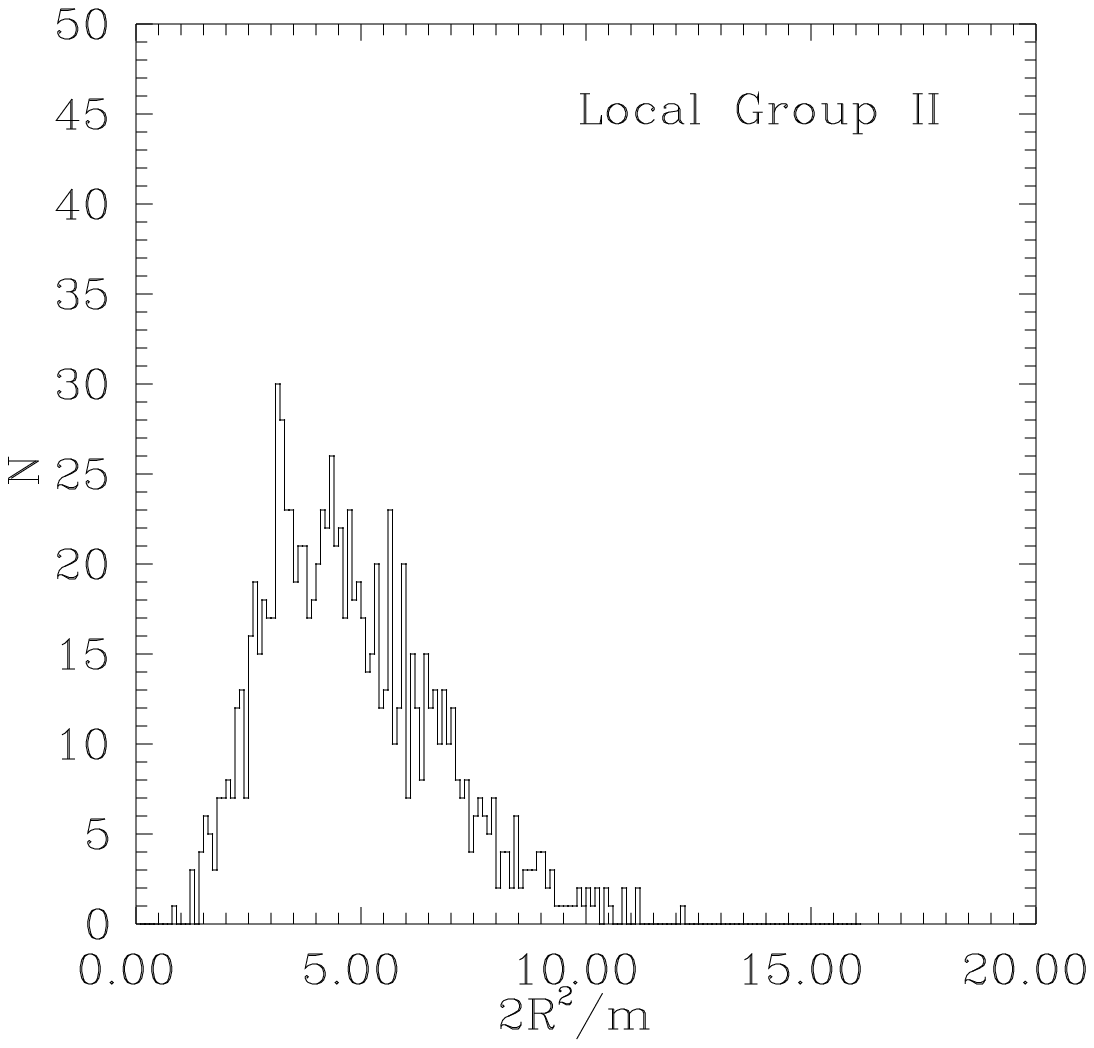}
\includegraphics{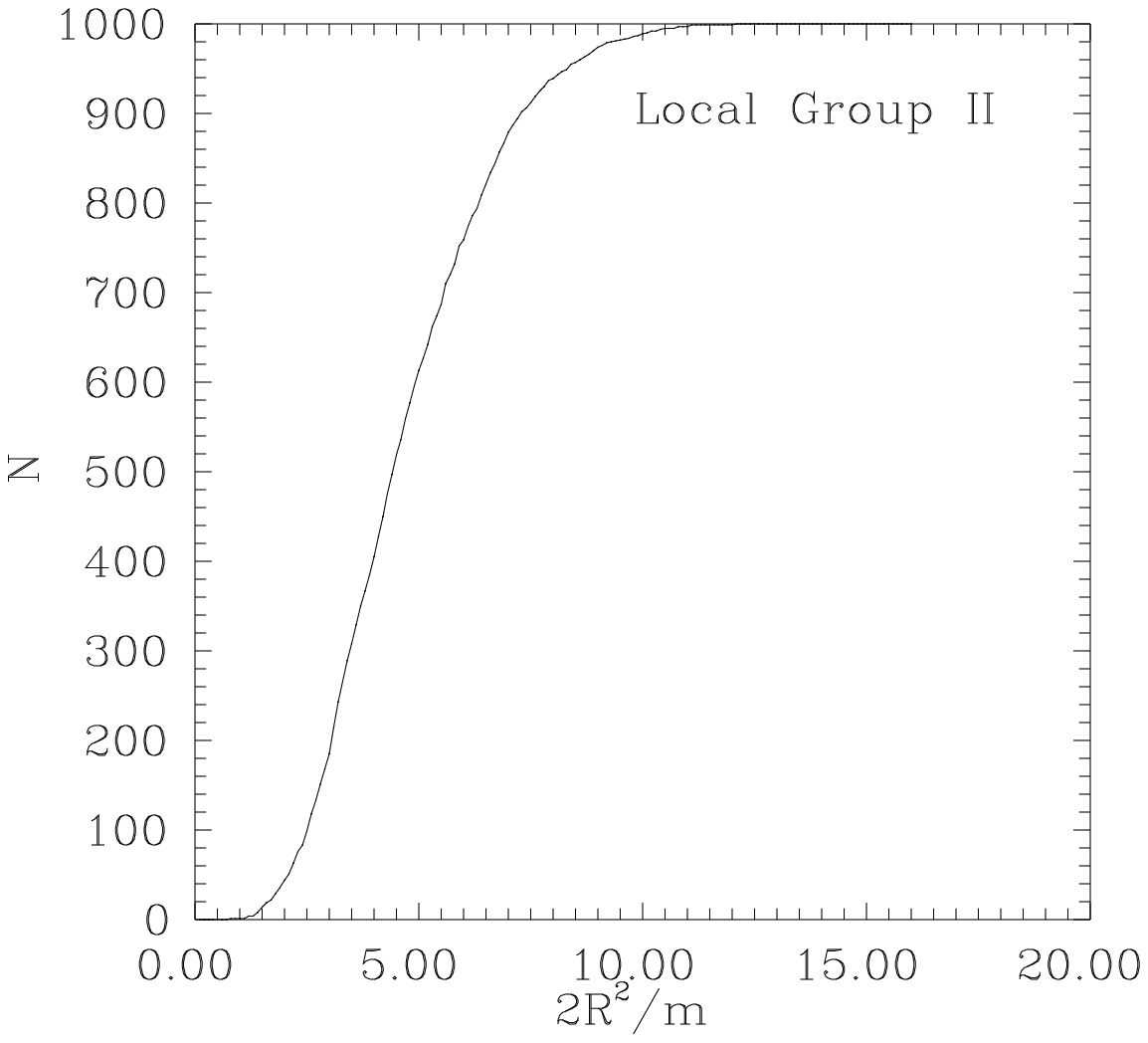}
\caption{Histogram for distribution of the height of the most significant peak
with data weighted using the Hann's function --- sample II (28 galaxies) (left
panel), and cumulated distribution of the height of the most significant peak
 with data  weighted using the Hann's function --- sample II (28 galaxies)
(right panel).\label{fig6}}
\end{figure*}
 
\begin{table*}
\begin{center}
\caption{List of galaxies.
\label{tbg}}
\footnotesize{
\begin{tabular}{crrrrrrrrrrr}
\hline\hline
$1$&$2$&$3$&$4$&$5$&$6$&$7$&$8$&$9$&$10$&$11$&$12$ \\
Nr  &PGC   &NGC   &name     &morf.      &$\alpha_{1950.0}$&$\delta_{1950.0}$&l&b&$v_{\mathrm{I}}$&$v_{\mathrm{B}}$& \\
\hline
  1 & 1305 &      & IC 10   & Ir IV     &$ 00 17 42$ &$ +59 00 52 $ &$ 118.97$&$  -03.34$ &$-344\pm 4 $&$-344\pm 5  $ &    \\
  2 &      &      & Cetus   & dE4       &$ 00 23 36$ &$ -11 19 00 $ &$ 101.40$&$  -72.80$ &$          $&$           $ & $N$\\
  3 & 2004 &  147 & DDO 3   & dE5       &$ 00 30 27$ &$ +48 13 56 $ &$ 119.82$&$  -14.25$ &$-193\pm 3 $&$-193\pm 3  $ &    \\
  4 & 2121 &      & And III & Dsph      &$ 00 32 42$ &$ +36 14 00 $ &$ 119.34$&$  -26.25$ &$-355\pm10 $&$-355\pm10  $ & $N$\\
  5 & 2329 &  185 &         & dE3p      &$ 00 36 12$ &$ +48 03 50 $ &$ 120.79$&$  -14.48$ &$-208\pm 7 $&$-202\pm 7  $ &    \\
  6 & 2555 &  221 & M32     & E2        &$ 00 36 58$ &$ +40 35 29 $ &$ 121.15$&$  -21.98$ &$-205\pm 3 $&$-205\pm 3  $ &    \\
  7 & 2429 &  205 &         & E6p       &$ 00 37 39$ &$ +41 24 44 $ &$ 120.72$&$  -21.14$ &$-239\pm 3 $&$-244\pm 3  $ &    \\
  8 & 2557 &  224 & M31     & Sb I-II   &$ 00 40 00$ &$ +40 59 43 $ &$ 121.17$&$  -21.57$ &$-299\pm 1 $&$-301\pm 1  $ &    \\
  9 & 2666 &      & And I   & Dsph      &$ 00 42 48$ &$ +37 46 00 $ &$ 121.65$&$  -24.82$ &$-380\pm 2 $&$-380\pm 2  $ & $N$\\
 10 & 3085 &      & SMC     & Ir IV-V   &$ 00 50 53$ &$ -73 04 18 $ &$ 302.81$&$  -44.33$ &$ 163\pm 4 $&$ 148\pm 4  $ &    \\
 11 & 3589 &      & Scl     & Dsph      &$ 00 57 47$ &$ -33 58 42 $ &$ 287.53$&$  -83.16$ &$ 107\pm 3 $&$ 110\pm 3  $ &    \\
 12 & 3792 &      & LGS 3   & DIr       &$ 01 01 12$ &$ +21 37 00 $ &$ 126.75$&$  -40.89$ &$-277\pm 5 $&$-286\pm 5  $ &    \\
 13 & 3844 &      & IC 1613 & Ir V      &$ 01 02 20$ &$ +01 51 56 $ &$ 129.79$&$  -60.56$ &$-236\pm 1 $&$-232\pm 1  $ &    \\
 14 & 4126 &  404 &         & S0        &$ 01 06 39$ &$ +35 27 06 $ &$ 127.03$&$  -27.01$ &$ -45\pm 9 $&$           $ & $M$\\
 15 &      &      & AND V   & dE        &$ 01 07 18$ &$ +47 22 00 $ &$ 126.20$&$  -15.10$ &$-403\pm 4 $&$-403\pm 4  $ & $N$\\
 16 & 4601 &      & And II  & Dsph      &$ 01 13 36$ &$ +33 11 00 $ &$ 128.89$&$  -29.14$ &$-188\pm 3 $&$-188\pm 3  $ & $N$\\
 17 & 5818 &  598 & M33     & Sc II-III &$ 01 31 02$ &$ +30 24 15 $ &$ 133.61$&$  -31.33$ &$-180\pm 1 $&$-181\pm 1  $ &    \\
 18 &      &      & Phoenix & Ir        &$ 01 49 00$ &$ -44 42 00 $ &$ 272.19$&$  -68.95$ &$  56\pm 3 $&$  56\pm 6  $ &    \\
 19 & 9892 &      & Maffei 1& E         &$ 02 32 36$ &$ +59 26 00 $ &$ 135.83$&$  -00.57$ &$   2\pm72 $&$           $ & $M$\\
 20 & 10093&      & Fornax  & Dsph      &$ 02 37 55$ &$ -34 39 48 $ &$ 237.10$&$  -65.65$ &$  53\pm 3 $&$  53\pm 2  $ &    \\
 21 & 10217&      & Maffei 2& Sbc       &$ 02 38 08$ &$ +59 23 24 $ &$ 136.50$&$  -00.33$ &$  -1\pm 6 $&$           $ & $M$\\
 22 & 15345& 1569 &         & Ir        &$ 04 26 06$ &$ +64 44 18 $ &$ 143.68$&$  +11.24$ &$ -77\pm 6 $&$           $ &    \\
 23 & 15488& 1560 &         & Sd        &$ 04 27 08$ &$ +71 46 29 $ &$ 138.37$&$  +16.02$ &$ -40\pm 7 $&$           $ &    \\
 24 & 15439&      & UGCA 92 & Ir        &$ 04 27 24$ &$ +63 30 00 $ &$ 144.71$&$  +10.51$ &$-105\pm 5 $&$           $ &    \\
 25 &      &      & Cam A   & Ir        &$ 04 31 30$ &$ +71 25 00 $ &$ 138.88$&$  +16.04$ &$-127\pm ?:$&$           $ & $N$\\
 26 & 17223&      & LMC     & Ir III-IV &$ 05 24 00$ &$ -69 48 00 $ &$ 280.47$&$  -32.89$ &$ 272\pm 8 $&$ 275\pm 2  $ &    \\
 27 & 19441&      & Carina  & Dsph      &$ 06 40 24$ &$ -50 55 00 $ &$ 260.11$&$  -22.22$ &$ 223\pm 3 $&$ 223\pm 3  $ &    \\
 28 &      &      & ARGO    & Ir        &$ 07 04 30$ &$ -58 27 00 $ &$ 268.96$&$  -21.15$ &$ 554\pm10 $&$           $ &
\end{tabular}
}
\end{center}
\end{table*}
 
\begin{table*}
\begin{center}
\caption{List of galaxies.
\label{tbh}}
\footnotesize{
\begin{tabular}{crrrrrrrrrrr}
\hline\hline
$1$&$2$&$3$&$4$&$5$&$6$&$7$&$8$&$9$&$10$&$11$&$12$ \\
Nr  &PGC   &NGC   &name     &morf.      &$\alpha_{1950.0}$&$\delta_{1950.0}$&l&b&$v_{\mathrm{I}}$&$v_{\mathrm{B}}$& \\
\hline
 29 & 21600&      & DDO 47  & Ir        &$ 07 39 00$ &$ +16 55 00 $ &$ 203.10$&$  +18.54$ &$ 270\pm 4 $&$           $ & $M$\\
 30 & 28868&      & Leo A   & Ir V      &$ 09 56 24$ &$ +30 59 00 $ &$ 196.90$&$  +52.41$ &$  26\pm 2 $&$  24\pm 4  $ &    \\
 31 & 28913&      &Sextans B& Ir        &$ 09 57 23$ &$ +05 34 22 $ &$ 233.20$&$  +43.78$ &$ 301\pm 2 $&$           $ &    \\
 32 & 29128& 3109 & DDO 236 & Ir        &$ 10 00 49$ &$ -25 55 00 $ &$ 262.10$&$  +23.07$ &$ 403\pm 1 $&$           $ &    \\
 33 &      &      & Antlia  & dE3       &$ 10 01 48$ &$ -27 05 00 $ &$ 263.10$&$  +22.32$ &$ 361\pm 2 $&$           $ &    \\
 34 & 29488&      & Leo I   & Dsph      &$ 10 05 47$ &$ +12 33 10 $ &$ 225.98$&$  +49.11$ &$ 285\pm 2 $&$ 287\pm 5  $ &    \\
 35 & 29653&      &Sextans A& Ir        &$ 10 08 30$ &$ -04 28 00 $ &$ 246.17$&$  +39.86$ &$ 325\pm 3 $&$           $ &    \\
 36 &      &      & Sextans & dE4       &$ 10 10 42$ &$ -01 24 00 $ &$ 243.55$&$  +42.27$ &$ 224\pm 2 $&$ 226\pm 1  $ &    \\
 37 & 34176&      & Leo II  & Dsph      &$ 11 10 50$ &$ +22 25 32 $ &$ 220.17$&$  +67.23$ &$  76\pm 2 $&$  76\pm 1  $ &    \\
 38 & 35286&      & UGC 6456& P         &$ 11 24 36$ &$ +79 16 00$ &$ 127.84$&$  +37.33$ &$ -92\pm 5 $&$           $ & $M$\\
 39 & 39346& 4236 &         & Sdm       &$ 12 14 22$ &$ +69 44 36 $ &$ 127.43$&$  +47.36$ &$   0\pm 4 $&$           $ & $M$\\
 40 & 44491&      & Gr 8    & Ir        &$ 12 56 06$ &$ +14 29 00 $ &$ 310.72$&$  +76.98$ &$ 216\pm 3 $&$           $ &    \\
 41 & 50961&      & DDO 187 & Ir        &$ 14 13 36$ &$ +23 17 00 $ &$  25.57$&$  +70.47$ &$ 154\pm 4 $&$           $ & $M$\\
 42 & 54074&      & UMi     & Dsph      &$ 15 08 12$ &$ +67 23 00 $ &$ 104.97$&$  +44.84$ &$-250\pm 2 $&$-247\pm 1  $ &    \\
 43 & 60095&      & Draco   & Dsph      &$ 17 19 24$ &$ +57 57 50 $ &$  86.37$&$  +34.72$ &$-289\pm 2 $&$-293\pm 1  $ &    \\
 44 &      &      & Galaxy  & Sbc       &$ 17 42 24$ &$ -28 55 50 $ &$   0.00$&$  +00.00$ &$   0\pm10 $&$  16\pm 0  $ &    \\
 45 &      &      & Sagittar& dE7       &$ 18 51 54$ &$ -30 30 00 $ &$   5.65$&$  -14.08$ &$ 140\pm 5 $&$ 142\pm 1  $ &    \\
 46 & 63287&      & Sgr     & Ir        &$ 19 27 06$ &$ -17 47 00 $ &$  21.06$&$  -16.29$ &$ -79\pm 2 $&$ -79\pm 4  $ &    \\
 47 & 63613& 6822 & DDO 209 & Ir IV-V   &$ 19 42 08$ &$ -14 55 29 $ &$  25.34$&$  -18.40$ &$ -49\pm 6 $&$ -56\pm 2  $ &    \\
 48 & 65367&      & Aqr     & Ir        &$ 20 44 06$ &$ -13 02 00 $ &$  34.05$&$  -31.35$ &$-131\pm 5 $&$-131\pm 5  $ &    \\
 49 & 67908&      & IC 5152 & Ir IV     &$ 21 59 26$ &$ -51 32 18 $ &$ 343.92$&$  -50.19$ &$ 121\pm 4 $&$           $ &    \\
 50 &      &      & Tucana  & dE5       &$ 22 38 30$ &$ -64 41 00 $ &$ 322.90$&$  -47.37$ &$ 130\pm 2 $&$ 130\pm 2  $ & $N$\\
 51 &      &      & UKS2323-& Ir        &$ 23 23 48$ &$ -32 40 00 $ &$  11.86$&$  -70.86$ &$  62\pm 6 $&$           $ &    \\
 52 &      &      & AND VII & dE3       &$ 23 24 12$ &$ +30 25 00 $ &$ 109.50$&$  -09.90$ &$-307\pm 2 $&$-307\pm 2  $ & $N$\\
 53 & 71538&      & Peg Ir  & Ir V      &$ 23 26 03$ &$ +14 28 16 $ &$  94.77$&$  -43.55$ &$-181\pm 2 $&$-182\pm 2  $ &    \\
 54 &      &      & AND VI  & dE3       &$ 23 49 12$ &$ +24 18 00 $ &$ 106.00$&$  -36.30$ &$-354\pm 3 $&$-354\pm 3  $ & $N$\\
 55 &  143 &      & WLM     & Ir IV-V   &$ 23 59 23$ &$ -15 43 43 $ &$  75.87$&$  -73.61$ &$-116\pm 2 $&$-116\pm 2  $ &    \\
\hline
\end{tabular}
}
\end{center}
\end{table*}

\begin{table}
\begin{center}
\caption{Motion of the Sun.
\label{tbs}}
\begin{tabular}{crrr}
\hline\hline
Sample&$V_{\mathrm{h}}$&$l$&$b$\\
\hline
 $a$&$232 $&$ 88 $&$  2  $\\
 $b$&$233 $&$ 93 $&$  2  $\\
 $c$&$213 $&$ 93 $&$  2  $\\
 $d$&$234 $&$ 98 $&$  2  $\\
 $e$&$  0 $&$  0 $&$  0  $\\
 $f$&$306 $&$ 99 $&$ -3.5$\\
 $g$&$305 $&$ 94 $&$-34  $\\
 $h$&$308 $&$105 $&$ -7  $\\
 $i$&$224 $&$109 $&$ 65  $\\
 $j$&$316 $&$ 93 $&$ -4  $\\
\hline
\end{tabular}
\end{center}
\end{table}
 
\begin{table}
\begin{center}
\caption{Statistical analysis of samples investigated in
the  Local Group (Sample I) \label{tbk}}
\begin{tabular}{crrrr}
\hline\hline
Sample&$V$&$P(s)$&$SI$&$Q$\\
\hline
 $1a$&$ 23.8 $&$ 0.232 $&$ 139.4 $&$ 1.394 $\\
 $1b$&$ 25.7 $&$ 0.151 $&$ 135.9 $&$ 1.359 $\\
 $1c$&$ 31.2 $&$ 0.351 $&$ 131.2 $&$ 1.312 $\\
 $1d$&$ 25.6 $&$ 0.477 $&$ 113.8 $&$ 1.138 $\\
 $1e$&$ 26.8 $&$ 0.081 $&$ 110.3 $&$ 1.103 $\\
 $1f$&$ 37.0 $&$ 0.350 $&$  88.6 $&$  .886 $\\
 $1g$&$ 37.2 $&$ 0.561 $&$ 108.1 $&$ 1.081 $\\
 $1h$&$ 56.5 $&$ 0.922 $&$ 101.1 $&$ 1.011 $\\
 $1i$&$ 36.2 $&$ 0.439 $&$ 113.6 $&$ 1.136 $\\
 $1j$&$ 41.5 $&$ 0.809 $&$ 109.9 $&$ 1.099 $\\
\hline
\end{tabular}
\end{center}
\end{table}
 
\begin{table*}
\begin{center}
\caption{Statistical analysis of samples investigated in
the  Local Group (Samples II - V) \label{tbl}}
\begin{tabular}{crrrr|crrrr}
\hline\hline
Sample&$V$&$P(s)$&$SI$&$Q$&Sample&$V$&$P(s)$&$SI$&$Q$\\
\hline
 $2a$&$ 33.3 $&$ 0.219 $&$ 143.1 $&$ 1.431 $&$4a$&$ 28.4 $&$ 0.054 $&$ 141.9 $&$ 1.419 $\\
 $2b$&$ 36.8 $&$ 0.229 $&$ 142.2 $&$ 1.422 $&$4b$&$ 28.4 $&$ 0.076 $&$ 118.2 $&$ 1.182 $\\
 $2c$&$ 20.6 $&$ 0.508 $&$ 115.3 $&$ 1.153 $&$4c$&$ 31.3 $&$ 0.323 $&$ 126.9 $&$ 1.269 $\\
 $2d$&$ 37.5 $&$ 0.907 $&$  99.8 $&$  .998 $&$4d$&$ 20.1 $&$ 0.584 $&$ 105.0 $&$ 1.050 $\\
 $2e$&$ 51.8 $&$ 0.363 $&$  98.7 $&$  .987 $&$4e$&$ 27.2 $&$ 0.154 $&$  95.0 $&$  .950 $\\
 $2f$&$ 56.3 $&$ 0.595 $&$ 103.8 $&$ 1.038 $&$4f$&$ 37.2 $&$ 0.354 $&$ 107.9 $&$ 1.079 $\\
 $2g$&$ 43.3 $&$ 0.125 $&$  95.5 $&$  .955 $&$4g$&$ 37.8 $&$ 0.472 $&$ 114.3 $&$ 1.143 $\\
 $2h$&$ 55.7 $&$ 0.595 $&$  96.4 $&$  .964 $&$4h$&$ 52.6 $&$ 0.996 $&$ 104.2 $&$ 1.042 $\\
 $2i$&$ 31.2 $&$ 0.235 $&$ 115.4 $&$ 1.154 $&$4i$&$ 29.6 $&$ 0.053 $&$ 114.3 $&$ 1.143 $\\
 $2j$&$ 20.0 $&$ 0.482 $&$  95.7 $&$  .957 $&$4j$&$ 25.7 $&$ 0.217 $&$ 115.4 $&$ 1.154 $\\
 $3a$&$ 23.6 $&$ 0.156 $&$ 147.4 $&$ 1.474 $&$5a$&$ 23.5 $&$ 0.558 $&$ 130.0 $&$ 1.300 $\\
 $3b$&$ 36.8 $&$ 0.241 $&$ 135.7 $&$ 1.357 $&$5b$&$ 27.6 $&$ 0.159 $&$ 119.8 $&$ 1.198 $\\
 $3c$&$ 23.2 $&$ 0.203 $&$ 145.3 $&$ 1.453 $&$5c$&$ 20.4 $&$ 0.375 $&$ 105.5 $&$ 1.055 $\\
 $3d$&$ 25.4 $&$ 0.697 $&$ 106.5 $&$ 1.065 $&$5d$&$ 29.9 $&$ 0.766 $&$  99.0 $&$  .990 $\\
 $3e$&$ 26.9 $&$ 0.100 $&$  88.5 $&$  .885 $&$5e$&$ 57.8 $&$ 0.168 $&$  91.3 $&$  .913 $\\
 $3f$&$ 48.0 $&$ 0.135 $&$ 103.0 $&$ 1.030 $&$5f$&$ 20.5 $&$ 0.517 $&$ 104.5 $&$ 1.045 $\\
 $3g$&$ 37.3 $&$ 0.370 $&$ 104.9 $&$ 1.049 $&$5g$&$ 43.7 $&$ 0.105 $&$  91.4 $&$  .914 $\\
 $3h$&$ 52.5 $&$ 0.885 $&$ 107.1 $&$ 1.071 $&$5h$&$ 55.7 $&$ 0.839 $&$ 100.2 $&$ 1.002 $\\
 $3i$&$ 29.7 $&$ 0.075 $&$ 116.8 $&$ 1.168 $&$5i$&$ 47.6 $&$ 0.588 $&$ 115.7 $&$ 1.157 $\\
 $3j$&$ 25.6 $&$ 0.809 $&$ 108.5 $&$ 1.085 $&$5j$&$ 20.0 $&$ 0.548 $&$  94.4 $&$  .944 $\\
\hline
\end{tabular}
\end{center}
\end{table*}
 
\clearpage
 
\begin{table}
\begin{center}
\caption{Statistical analysis of samples investigated in
the  Local Group (with data weighted using the Hann's function). \label{tbn}}
\begin{tabular}{crrrr}
\hline\hline
Sample&$V$&$P(s)$&$SI$&$Q$\\
\hline
 $2a$&$ 32.9 $&$ 0.353 $&$ 138.1 $&$ 1.381 $\\
 $2b$&$ 36.9 $&$ 0.367 $&$  98.3 $&$  .983 $\\
 $2c$&$ 20.6 $&$ 0.195 $&$ 116.7 $&$ 1.167 $\\
 $2d$&$ 33.5 $&$ 0.802 $&$  75.9 $&$  .759 $\\
 $2e$&$ 51.9 $&$ 0.230 $&$  96.4 $&$  .964 $\\
 $2f$&$ 30.5 $&$ 0.985 $&$  67.7 $&$  .677 $\\
 $2g$&$ 43.4 $&$ 0.341 $&$  70.2 $&$  .702 $\\
 $2h$&$ 77.8 $&$ 0.913 $&$  90.8 $&$  .908 $\\
 $2i$&$ 31.8 $&$ 0.551 $&$  86.9 $&$  .869 $\\
 $2j$&$ 35.1 $&$ 0.943 $&$  66.4 $&$  .664 $\\
 $3a$&$ 81.3 $&$ 0.697 $&$  73.9 $&$  .739 $\\
 $3b$&$100.0 $&$ 0.990 $&$  67.5 $&$  .675 $\\
 $3c$&$ 84.7 $&$ 0.968 $&$  75.1 $&$  .751 $\\
 $3d$&$ 23.1 $&$ 0.492 $&$  85.3 $&$  .853 $\\
 $4a$&$ 28.5 $&$ 0.639 $&$  80.9 $&$  .809 $\\
 $4b$&$ 20.4 $&$ 0.768 $&$  68.3 $&$  .683 $\\
 $4c$&$ 20.1 $&$ 0.816 $&$  75.8 $&$  .758 $\\
 $4d$&$ 22.9 $&$ 0.733 $&$  71.6 $&$  .716 $\\
 $5a$&$ 27.4 $&$ 0.257 $&$ 107.3 $&$ 1.073 $\\
 $5b$&$ 36.5 $&$ 0.729 $&$  87.3 $&$  .873 $\\
 $5c$&$ 20.4 $&$ 0.219 $&$ 101.5 $&$ 1.015 $\\
 $5d$&$ 21.3 $&$ 0.834 $&$  82.0 $&$  .820 $\\
\hline
\end{tabular}
\end{center}
\end{table}
 
\end{document}